# Transit Timing Variations of Five Transiting Planets

Ö. Baştürk [1, a], E.M. Esmer [1], Ş. Torun [1], S. Yalçınkaya [1], F. El Helweh [2],
E. Karamanlı [2], M. Öncü [2], H.Ö. Albayrak [2], F.A.M. Akram [2], M.G. Kahraman [2],
S. Sufi [2], M. Üzümcü [1], and F. Davoudi [3]

[1]*Ankara University, Faculty of Science, Astronomy & Space Sciences Dept., TR-06100 Ankara / Turkey*
[2]*Bilkent University, Science Faculty, Physics Department, TR-06800 Ankara / Turkey*
[3]*Department of Physics, University of Zanjan, P.O. Box 45195-313, Zanjan, Iran*

[a] Corresponding author: obasturk@ankara.edu.tr

**Abstract.** 4116 exoplanets have been discovered so far in 3062 planetary systems in total, 669 of which are multi-planet systems (http://exoplanet.eu, 2019-09-14). Finding a large, short orbital period planet transiting its star is a relatively easy task thanks to space born and ground based transit surveys. Whether they are the sole planets orbiting their host stars or they have companion planets is a worthy question to ask, because an average of ~1.34 planets per star is not a number expected from planet formation theories and a comparison with our own Solar System hosting at least 8 planets and many more bodies in planetary mass limits Transiting planets provide a unique opportunity to search for unseen additional bodies gravitationally bound to the system, which doesn't have to transit the host star. It is possible to detect the motion of the center of mass of the observed transiting planet host star duo due to the gravitational tugs of the unseen bodies from the Roemer delay. In order to achieve the goal, determination of the mid times of the transits of the planets in high precision and accuracy is a primary condition. These mid transit times have to be detrended from the orbital motion of the Earth about the Sun, which brings its own delay because it also changes the distance between the observed planet system and the Earth. Then potential periodic variations in the transit timings due to the so-called Light Travel Time Effect (LiTE) is searched. We present transit timing variations and update the ephemeris information of 5 transiting planets; HAT P 23b, WASP 103b, GJ 1214b, WASP 69b, and KELT 3b within this contribution. We have collected all the quality transit light curves of these exoplanets from the literature and observations of amateur astronomers made available through Exoplanet Transit Database (ETD), converted them to Dynamic Barycentric Julian days (BJD-TDB), constituted their Transit Timing Variation (TTV) diagrams, and updated their ephemeris information based on Markov Chain Monte Carlo (MCMC) analyses. Finally, we have carried out frequency analyses for all the planets in our sample and present the results.

## Introduction

Observations of exoplanets during their transits in front of their host stars have provided wealthy of information to astronomers since the first attempts at the beginning of this century [8]. Almost three thousand planets have been discovered thanks to their transits in over two thousand planetary systems. Important parameters such as their radii, orbital sizes, shapes, inclinations and periods and their equivalent temperatures are determined with some help from established theory of stellar astrophysics and complementary radial velocity observations. Transit timings (ingress, mid-transit, egress) are observational parameters determined within the analysis of a transit light curve as well as its shape and its depth. In the presence of gravitationally bound, yet unseen bodies, the timings of a transit vary due to the perturbation of the orbit of the transiting body and / or the Roemer delay caused by the reflex motion of the observed system in transit about the common center of mass with the unseen body. Therefore, these variations help us find the unseen planets as well as determine their dynamic properties. It also helps to lift-off the mass-radius degeneracy for both planets in the system if the third body is also caught during its transits [17]. An important class of transiting planets is hot-Jupiters, so-called because they are gas giants orbiting stars in short-period orbits. Since the transit technique is biased towards finding such planets, a significant fraction of all the transiting planets are hot-Jupiters although their unbiased occurrence rates should be much smaller. There have been only a few additional planet discoveries around stars hosting hot-Jupiters. Whether they are the only planets orbiting their host stars or not is an important question therefore; answer to which will have important consequences on planet formation and migration theories. Hence, observations of transit timing variations in systems bearing hot-Jupiters is crucial also as inputs to formation and dynamics of planetary systems.

With these motivations, we have started a project following hot-Jupiter bearing exoplanet systems observationally. We have carried out transit observations of a number of exoplanet systems selected according to the radial velocity residuals that they display and eccentricities of their orbits. We have made use of observational facilities in Turkey, such as 1 m Turkish telescope T100, located in the Bakırlıtepe campus of TÜBİTAK National Observatory of Turkey (TUG) and 35 cm telescope T35 in Ankara University Kreiken Observatory (AUKR). Within this contribution, we present the preliminary analyses of mid-transit timings that we determined from our own observations of 5 selected exoplanets from our target list, together with the mid-transit times that we collected from the literature as well as that from the observations of amateur astronomers presented in the Exoplanet Transit Database (ETD, http://var2.astro.cz/ETD/) [16]. We have updated the ephemeris information (reference epoch $T_0$, and orbital period P) based on our analyses of these heterogeneous data sets of mid-transit times as a result. We have also performed Fourier analyses to search for periodicities in the residuals from the linear fits to the data. But we have not found any statistically significant peak in our power spectra for any of the analysed planetary systems.

## Observations and Data Acquisition

We observed 6 transits of our five targets with T100 and 2 transits of WASP-103b and WASP-69b with T35 in good atmospheric conditions. T100 has a cryogenically cooled, SI-1100 CCD camera with 4098 x 4098 pixels attached on it and it is located 2500 m above sea level, in close proximity of the city of Antalya in the south coast of Turkey. T35 is a 35 cm telescope located in the campus of Ankara University Kreiken Observatory in Ankara. It has a 1 megapixel Apogee ALTA U47+ CCD camera attached on it. We used a Bessel R filter except for the observation of WASP-69b transit on 2017-08-26 in Bessel I passband. We aggressively defocused the telescope in the observations with T100 in order to extend the exposure time to achieve better photometric precision, which had been shown to improve the timing precision [6, 18]. Although we have attempted multiple times with T100, we were not able to observe a transit of GJ-1214b [7], which is too faint ($m_V$ = 13.4) to observe with T35. We present a log of our observations in Table-1.

**TABLE 1.** Log of our observations with T100 and T35.

| Planet | Date of Obs. | Telescope | Band | $\sigma_{ph}$ [mmag] | Exp. Time [s] |
|---|---|---|---|---|---|
| HAT-P-23b | 2014-09-25 | T100 | Bessel R | 0.13 | 135 |
| WASP-103b | 2017-06-11 | T100 | Bessel R | 0.76 | 90 |
| WASP-103b | 2014-05-30 | T100 | Bessel R | 0.31 | 90 |
| WASP-103b | 2018-07-01 | T35 | Bessel R | 2.51 | 60 |
| WASP-69b | 2019-07-27 | T35 | Bessel R | 3.38 | 90 |
| WASP-69b | 2017-08-26 | T100 | Bessel I | 0.34 | 100 |
| WASP-69b | 2016-10-09 | T100 | Bessel R | 2.34 | 90 |
| KELT-3b | 2014-02-18 | T100 | Bessel R | 0.49 | 150 |

We made use of AstroImageJ (from hereafter AIJ) software package [10] for the reductions of our own CCD images (bias, dark, and flat corrections) as well as ensemble aperture photometry and airmass detrending. We selected best comparison stars for differential photometry in terms of spatial proximity on the detector, brightness, and spectral type, and averaged their fluxes in units of Analog-to-Digital Units (ADUs) to create high Signal-to-Noise Ratio (from hereafter SNR) artificial comparison stars with AIJ. We have employed various aperture sizes fixed for each of the nights of our observations, which heavily depend on the average atmospheric seeing throughout the night, to obtain the light curves that we detrended from the changing airmass effect at the end. We provide a T100 transit light curve of WASP-69b in Figure-1 and the T35 light curve of the same planet in Figure-2 in order to give the reader an idea about the precision of our observations, the level of which has been indicated by the nightly average of the photometric errors ($\sigma_{ph}$) measured from the target star with the AIJ in Table-1.

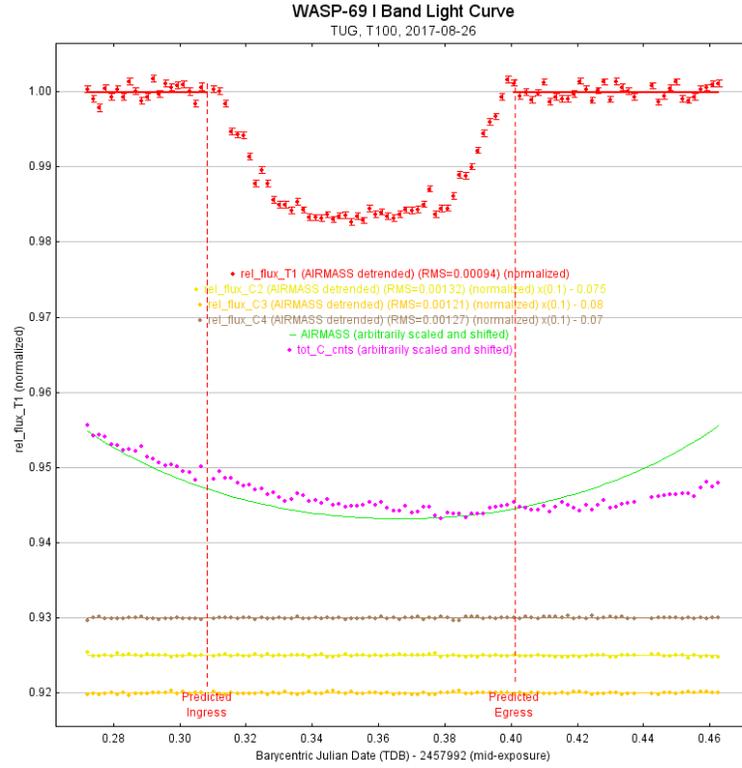

**FIGURE 1.** T100 transit light curve of WASP-69b on 2017-08-26 in Bessel I band.

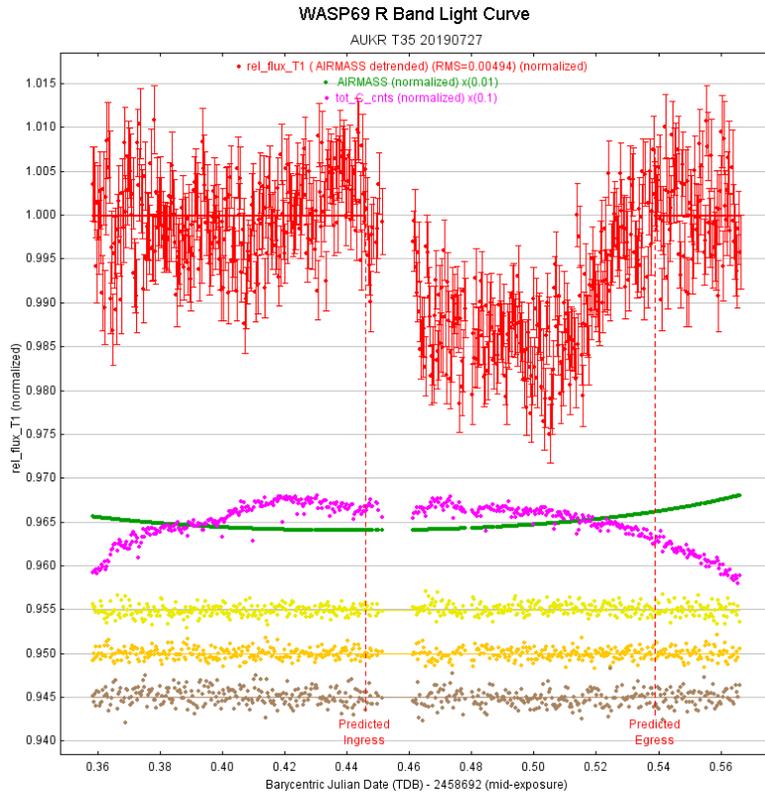

**FIGURE 2.** T35 transit light curve of WASP-69b on 2019-07-27 in Bessel R band.

We then modeled our light curves with the web-interface version of the software package EXOFAST [11] for speed. This version of EXOFAST (http://astroutils.astronomy.ohio-state.edu/exofast/exofast.shtml) makes use of Levenberg-Marquardt algorithm to fit a physical model by Mandel & Agol [13] to the observed transit data with the non-linear least squares method. We fixed the atmospheric ($T_{eff}$, log g, [Fe / H]) and orbital parameters (e, ω, P, K, Vγ) to their values determined from high resolution spectroscopic observation of the host stars published in the studies announcing the discovery of each the planets (HAT-P-23b: Bakos et al. 2011 [4], WASP-103b: Gillon et al. 2014 [12], WASP-69b: Anderson et al. 2014 [1], KELT-3b: Pepper et al. 2013 [15]), while adjusting the radius ratio ($R_p / R_*$), orbital inclination (i) and the scaled orbital size ($a / R_*$). We have acquired and fixed the coefficients of the quadratic limb darkening law ($u_1$ and $u_2$) by a linear interpolation from the tables given by Claret & Bloemen [9] with help of the online tool provided by EXOFAST for the Johnson-Cousins $R_c$ and $I_c$ filters, which have similar transmission curves with that of the Bessel filters that we used during the observations, and by making use of the stellar properties ($T_{eff}$, log g, [Fe / H]) from the discovery studies. We present a sample EXOFAST model of WASP-69b in Figure-3. The transit mid-time (in BJD-TDB) and its uncertainty are amongst the output parameters of a transit model.

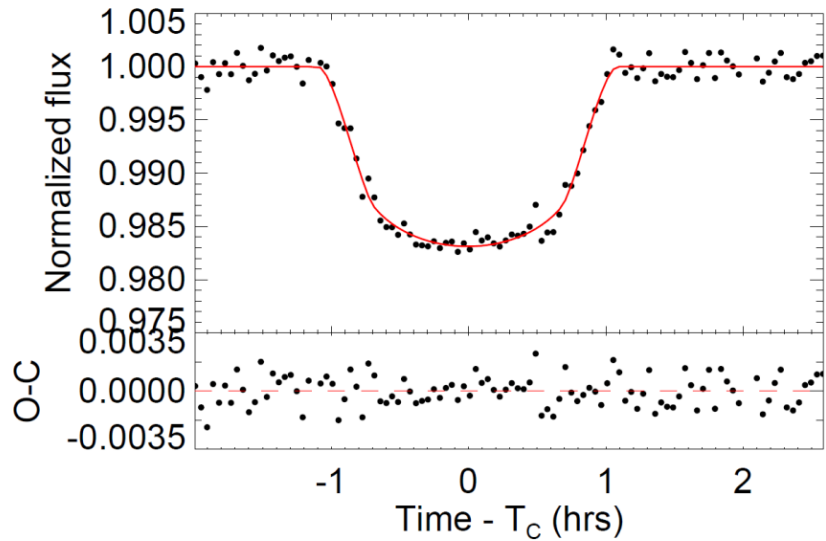

**FIGURE 3.** EXOFAST model (continuous curve on top panel) of the T100 transit light curve of WASP-69b (Figure-1) on 2017-08-26 in Bessel I band.

We have also collected all the mid-transit times published in the literature as well as that in the Exoplanet Transit Database (ETD), which publishes transit observations of amateur and professional observers online. We contacted the observers who submitted their light curves to ETD, in order to make sure whether the timings of their observations were synchronized with a Global Positioning System (GPS) device or an internet based time-server every few minutes. We also asked for the timing reference (JD, HJD, BJD-UTC, etc.), which they submitted their observations in. In the cases we were not sure about the timings, we asked for the raw data, reduced, performed differential aperture photometry with AIJ on the data, obtained the light curves and computed the mid-transit times in BJD-TDB and their uncertainties by ourselves and compared them with what the observers submitted to ETD, but used our own measurements in our analyses. We converted the mid-transit times that are not reported in BJD-TDB to this timing reference frame by using a code that we developed based on astropy functionality [2, 3].

## Analyses and Results

We calculated the mid-transit times, at which each observed transit had been predicted to be observed (denoted with the capital letter C) based on a reference epoch ($T_0$) and orbital period (P) that we took (and converted to BJD-TDB) from the papers announcing the discoveries of each of the planets and subtracted this value (C) from the times at which the transit was actually observed (denoted with the capital letter O), and hence computed the difference so-called O-C (observed minus calculated). We plotted these differences with respect to the number of cycles (epoch), thereby constructed the O-C (or TTV) diagrams. Since the reference epoch and orbital period have their own uncertainties, they expected to be accumulated over time causing a linear trend in the TTV diagrams. In order to search for periodicities in the TTV potentially caused by an unseen third planet, one has to remove such a trend from the O-C diagram.

In order to fit the linear trends on TTV diagrams, we took random samples from a set of epoch (dT) and orbital period (dP) values that we generated within a Markov Chain Monte Carlo (MCMC) by making use of Metropolis-Hastings algorithm. 500 walkers have been employed with 5000 iterations for each of the TTV diagrams in our MCMC runs, and the first 500 steps have been thrown away for the burn-in period. We used $X^2$ statistics as an indicator of the goodness of the fit, and integrated it into our log likelihood functions; and finally computed the posterior probability of each (dT,dP) pair based on these likelihood functions. The prior probability of each (dT,dP) pair has been the same (uniform priors). As a result, we obtained posterior probability distributions for the light elements ($T_0$ and P). We took the median values of each parameter, after leaving the values outside of $\pm 1\sigma$ from the distributions and accepted these as the best fitting parameters, that we list in Table-2 together with their uncertainties. We provide posterior probability distributions of ephemeris parameters together with the best fitting lines (red continuous lines) on TTV diagrams for each of the planets in Figures 4-8. Although the distributions are almost symmetric around the mean, there is a small asymmetry reflected in the asymmetric nature of errors in both directions. Despite being small, we have used more than necessary number of digits in the uncertainties of both parameters in Table-2 to indicate their asymmetric nature.

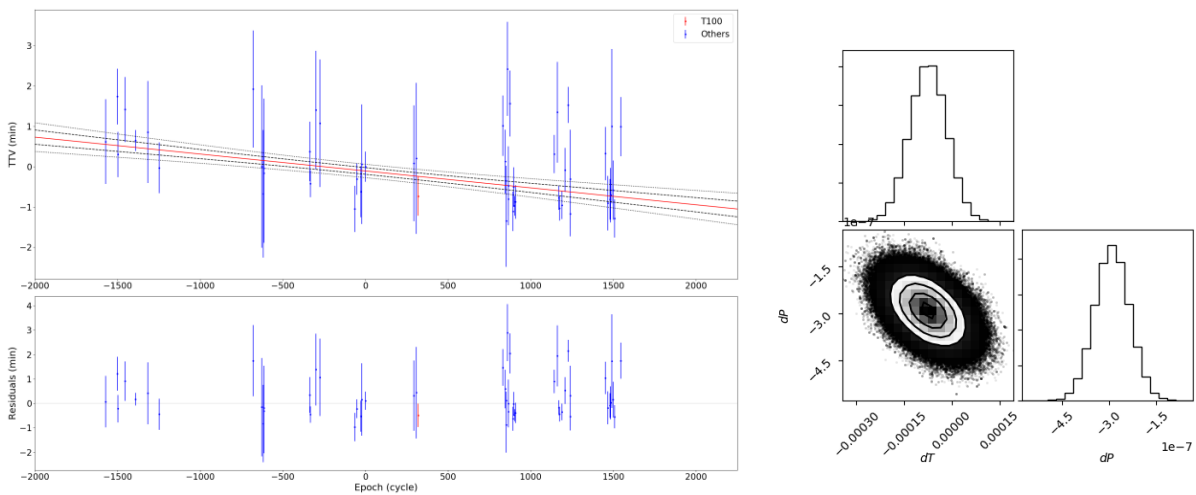

**FIGURE 4.** TTV diagram (upper panel on left) for HAT-P-23b and the residuals (lower panel left). Red data points are T100 observations, while the rest in blue are from the literature and ETD. Normalized posterior distributions for the light elements ($T_0$, P) (on right), and the correlation between those two parameters (lower left on right panel).

**TABLE 2.** Refined Ephemeris Information For 5 Transiting Exoplanets.

| Planet | $T_0$ [BJD-TDB] | $\sigma_T$ [s] | Period [days] | $\sigma_P$ [s] |
|---|---|---|---|---|
| HAT-P-23b | 2456539.390435 | +5.00 / -4.98 | 1.21288651 | +0.00471 / -0.00473 |
| WASP-103b | 2456459.599350 | +8.08 / -8.13 | 0.92554537 | +0.00739 / -0.00741 |
| WASP-69b | 2455748.834082 | +14.41 / -14.38 | 3.86814098 | +0.03641 / -0.03640 |
| KELT-3b | 2456023.480384 | +21.17 / -21.19 | 2.70339216 | +0.07490 / -0.07544 |
| GJ-1214b | 2454980.748826 | +2.72 / -2.72 | 1.58040428 | +0.06765 / -0.06750 |

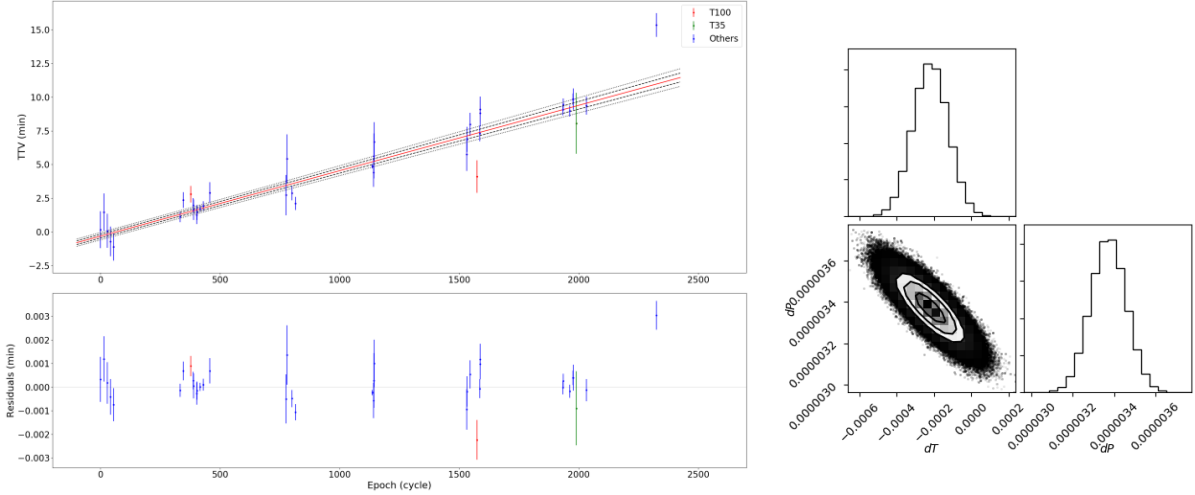

**FIGURE 5.** TTV diagram (upper panel on left) for WASP-103b and the residuals from the best fit (lower panel left). The same as Figure-4 except green data point shows the T35 observation.

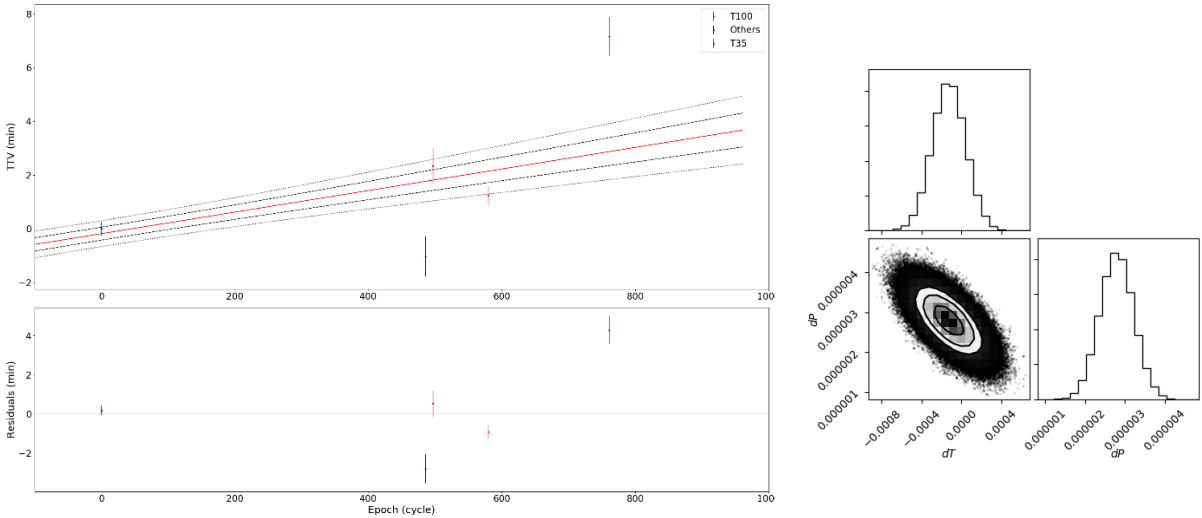

**FIGURE 6.** The same as Figıre-4 & 5 only for WASP-69b.

## Conclusions

We have collected all mid-transit times of 5 selected transiting exoplanets, observed and measured by amateur and professional observers. We carefully converted all these measurements to BJD-TDB, accounting for the locations of the observatories where these observations have been performed. We also observed at least one transit (except for GJ-1214b) of these planets with T100 and T35 telescopes located in TÜBİTAK National Observatory of Turkey (TUG) and Ankara University Kreiken Observatory (AUKR), respectively. We modeled our light curves with EXOFAST, and derived mid-transit times and their uncertainties with EXOFAST software package (Eastman 2017). Finally, we constructed so-called Transit Timing Variation (TTV) diagrams for all of our objects and computed linear ephemeris information within MCMC.

Our results show that the reference epoch can be determined very precisely within a level of only a couple of seconds (2.72 seconds for GJ-1214b is the largest uncertainty value). For all the planets, the observational errors are always less than a few minutes, allowing us to perform such a precise work. Linear model to TTVs is only successful to a level of a few minutes, and when they are detrended; the residuals display a variation from a few seconds (WASP-103b) to almost 15 minutes (KELT-3b). That is why, we attempted at performing Fourier analysis on the residuals. However, we failed in finding significant periodicities in any of the exoplanet systems that we studied. This should not mean that the potential of these systems in terms of harboring additional bodies can be neglected. But, in fact, it is crucial to continue the observations of these systems, since a potential third body might have an orbital period much longer compared to the time baseline of the available data (~ 13 years for HAT-P-23b).

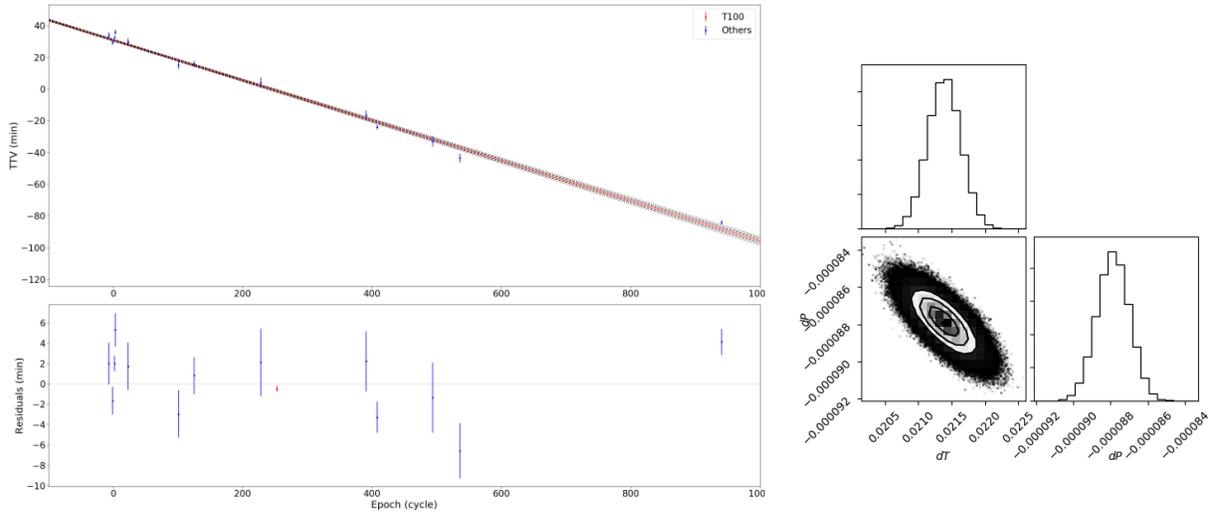

**FIGURE 7.** The same as Figure-4 only for KELT-3b.

Although the photometric precision we achieved with T100 thanks to the telescope defocusing technique and the optical quality and location of the telescope, resulted in one of the most precise timing measurements so far with less than a few second precision for all cases, smaller telescopes, like T35, can also provide sufficiently precise timing measurements to transit timing studies when they are effectively used in the observations of exoplanet transits. Therefore, we strongly suggest amateur and professional astronomers in small university observatories, who have access to 20 – 60 cm class telescopes, to observe transits of exoplanets, which they can plan by using the transit prediction tools in the online Exoplanet Transit Database. However, we would also like to stress that synchronization of the timings of observations with a GPS device or a time-server has utmost importance for both the precision and the accuracy of timing measurements required in TTV work.

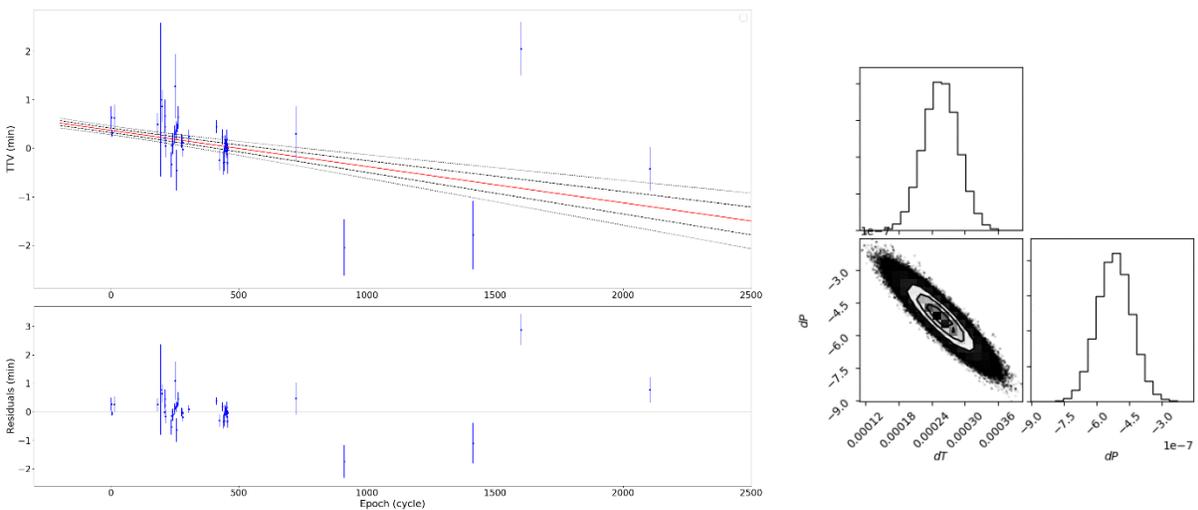

**FIGURE 8.** The same as Figure-4 except we have not been able to observe a transit of GJ-1214b within the telescope time available to us.

Finally, this work has made use of a heterogeneous set of data points collected from the literature and an open database. ETD users make use of the model by Pejcha [14] in order to determine the mid-transit times from their transit light curves. Their light curves are not detrended from the airmass effects in most cases. The mid-transit times from the literature have been determined by making use of various models and software packages. Therefore, we have started an effort to homogenize the data sets (as in [5]) that we use, starting with the detrending of the light curves and ending with the measurements of the mid-transit times and their uncertainty by ourselves with the same tools consistently. We think that comparison of the results of this contribution with that from the upcoming work will be helpful in terms of getting all transit observers prepare for future observations with similar standards, and researchers decide which data sets to use and which to discard in such a tedious work as the analysis of transit timing variations.

## ACKNOWLEDGMENTS

First and foremost, we thank Turkish Physical Society and all the organizers of 35[th] International Conference of Physics, held in beautiful Bodrum. We would also like to acknowledge the support by The Scientific and Technological Research Council of Turkey (TÜBİTAK) with the research project 118F042. We thank all the amateur and professional observers who let us use their transit observations in the Exoplanet Transit Database (ETD) and answered our never-ending questions about their setups. We also thank TÜBİTAK National Observatory of Turkey (TUG) with their partial support in using T100 for our transit observations with the projects numbered 12CT100-372, 16AT100-997, and 16BT100-1034. We appreciate the help before, during, and after the observations by TUG staff and night assistants. We are also grateful to all the student observers in Ankara University Kreiken Observatory (AUKR), which we would like to acknowledge for the observation time used for this project. Finally, we thank all the intern students from Bilkent University, who had their internships this summer in AUKR.